\documentclass[twocolumn,showpacs,amsmath,amssymb]{revtex4}
\usepackage{amsfonts}
\usepackage{graphicx}
\usepackage{epsfig}

\begin{document}
\title{Total suppression of superconductivity by high magnetic fields in YBa$_{2}
$Cu$_{3}$O$_{6.6}$}
\author{F. Rullier-Albenque$^{1}$, H. Alloul$^{2}$, Cyril Proust$^{3}$,
P. Lejay$^{4}$, A. Forget$^{1}$, D. Colson$^{1}$}

\affiliation{$^{1}$ SPEC, Orme des Merisiers, CEA, 91191 Gif sur
Yvette cedex, France\\
 $^{2}$ Laboratoire de Physique des Solides,
Univ. Paris-Sud, CNRS, UMR 8502, 91405 Orsay, France\\
$^{3}$Laboratoire National des Champs Magn\'{e}tiques Puls\'{e}s
(CNRS-UPS-INSA), Toulouse, France\\
$^{4}$CRTBT, CNRS, BP166X, 38042 Grenoble cedex, France}
\date{\today }

\begin{abstract}
We have studied in fields up to 60T the variation of the transverse
magnetoresistance (MR) of underdoped YBCO$_{6.6}$ crystals either
pure or with $T_{c}$ reduced down to 3.5K by electron irradiation.
We evidence that the normal state MR is restored above a threshold
field $H_{c}^{\prime}(T)$, which is found to vanish at
$T_{c}^{\prime}>>T_{c}$. In the pure YBCO$_{6.6}$ sample a 50 Tesla
field is already required to completely suppress the superconducting
fluctuations at $T_{c}$. While disorder does not depress the
pseudogap temperature, it reduces drastically the phase coherence
established at $T_{c}$, and weakly $H_{c}^{\prime}(0)$,
$T_{c}^{\prime}$ and the onset $T_{\nu}$ of the Nernst signal which
are more characteristic of the 2D local pairing.
\end{abstract}

\maketitle

Since the discovery of high temperature superconductivity the knowledge of the
normal state properties of the cuprates at $T<T_{c}$ has been somewhat hard to
explore since the upper critical fields of these materials are usually much
larger than any available experimental field. This has only been possible
using very high field facilities in systems with relatively low $T_{c}$ such
as La$_{2-x}$Sr$_{x}$CuO$_{4}$ (LSCO) \cite{Boebinger} or Bi$_{2}$Sr$_{2-x}%
$La$_{x}$CuO$_{6+\delta}$ (La-Bi2201) \cite{Ono}.

In fact, the determination of $H_{c2}$ in high-$T_{c}$ cuprates
remains a controversial issue. It was pointed out for long that
taking $H_{c2}(T)$ from the onset of the resistive transitions in
magnetic fields is questionable, due to the large liquid flux regime
above the melting line. This leads to a $T$ dependence of $H_{c2}$
with an upward curvature, similar to that found for the
irreversibility line deduced from the zero resistivity values
\cite{mackenzie,Osofsky}. Other authors have analysed the magnitude
of the magnetoconductance fluctuations near $T_{c}$ in a Ginzburg
Landau formalism to obtain the coherence length and then $H_{c2}$
\cite{Ando2, Bouquet}.

These approaches assume that $T_{c}$ is altogether the onset of the
amplitude of the superconducting order parameter and of phase
coherence, at variance with Emery and Kivelson's proposal
\cite{Emery} that only the phase coherence is broken at $T_{c}$
while the condensate amplitude remains finite. This suggestion has
gained strong support from high frequency conductivity measurements
of the short time scale phase coherence \cite{Corson}. Further
support has been provided by the observation above $T_{c}$ of a
large Nernst effect \cite{Wang-PRB1,RA3} combined with a diamagnetic
magnetization \cite{Wang-PRL2}. In this context the $H_{c2}$ line
determined as the field at which the Nernst and diamagnetic signals
reach zero \cite{Wang-science} does extend well above $T_{c}$.
However, presumably for sensitivity considerations, most of these
experiments have been performed below $T_{c}$.

In this letter, we propose a relatively simple method, based on an
analysis of the transverse magnetoresistance (MR) to determine the
$T$ dependence of the magnetic field $H_{c}^{^{\prime}}$ at which
the normal state transport is fully restored. We therefore provide
for the first time an experimental determination of the $(H,T)$
range of superconducting fluctuations, which extends up to
$T_{c}^{^{\prime}}>T_{c}$. This detection by transport properties of
the flux flow processes evidenced by Nernst measurements up to
$T_{\nu}\simeq T_{c}^{^{\prime}}$ therefore strongly validates the
vortex scenario. This is reinforced by the parabolic variation found
for $H_{c}^{^{\prime}}(T)$ which could thus be identified to the
microscopic depairing field as expected in the framework of the 2D
Kosterlitz Thouless transition \cite{Doniach}. Furthermore our data
have been taken on underdoped YBa$_{2}$Cu$_{3}$O$_{6.6}$ (YBCO)
which is found here to display a good metallic behavior, away from
the metal insulator crossover. This allows us altogether to perform
a study of the incidence of defects on $H_{c}^{^{\prime}}$,
$T_{c}^{^{\prime}}$ and $T_{\nu}$.

The samples used in this study have been described before, as well
as the electron irradiation procedure using the low temperature
facility of the Van der Graaff accelerator at the LSI (Ecole
Polytechnique, Palaiseau) \cite{RA3}. Pulsed magnetic fields have
been applied along the c axis in order to best suppress
superconductivity. Resistivity data have been taken initially on two
irradiated samples at the NHMFL in Los Alamos using short field
pulses (20ms), which implies ac measurements at high frequency
(250kHz). The other data have been taken at the LNCMP in Toulouse
with a long pulse (125ms) magnet and ac measurements at 50kHz.

The resistivity increase $\delta\rho=\rho(H,T)-\rho(0,T)$ induced by
the magnetic field $H$ is displayed in Fig.1a for the pure
YBCO$_{6.6}$ crystal for $60$K$<T<150$K. At high temperature in the
normal state, $\delta\rho$ increases as $H^{2}$ as usually observed
for the classical transverse magnetoresistance in low magnetic
field. The value of the MR coefficient defined as
$a_{trans}(T)=\delta\rho(T)/\rho(T,0)H^{2}$ is $\thickapprox
5$x$10^{-6}$ T$^{-2}$ at 150K in good agreement with that obtained
for $H\leqslant14$T \cite{Harris, Ando2}. This observation is good
evidence that the weak field regime $\omega_{c}\tau\ll1$ with
$\omega_{c}$ the cyclotron frequency and $\tau$ the scattering time,
applies up to $60$T. As $T$ decreases, the variation of $\delta\rho$
progressively deviates from the $H^{2}$ dependence. In order to
better capture this evolution, we have plotted $\delta\rho$ versus
$H^{2}$ in Fig.1b. One can see that the normal state $H^{2}$
dependence of $\delta\rho$ remains clearly defined at high field and
the MR\ curve departs from this limiting behaviour at a field value
$H_{c}^{^{\prime}}$ which increases steadily as $T$ decreases. This
is for us the indication that $H_{c}^{^{\prime}}$ is the field at
which the superconducting contribution to the conductivity is
suppressed.

\begin{figure}
\centering
\includegraphics[width=8.0cm]{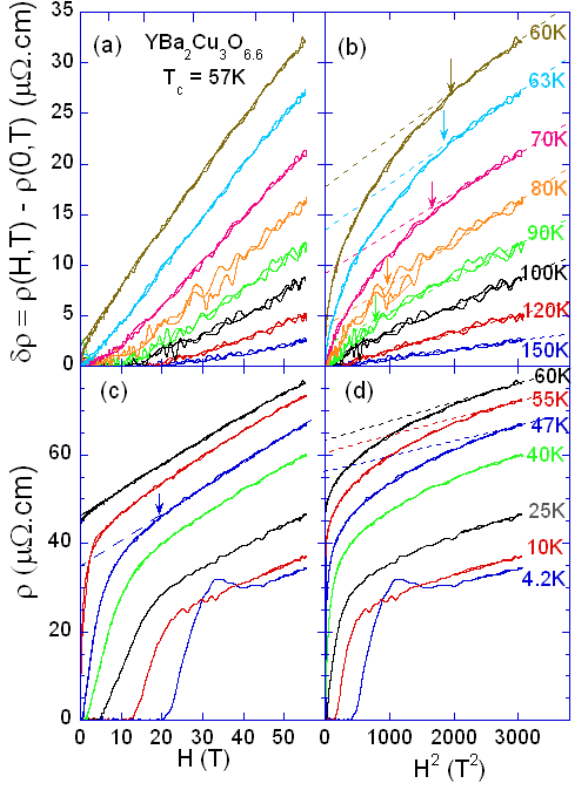}
\caption{(Color online) Magnetoresistance (above $T_{c})$ and
resistivity (below $T_{c})$ are plotted versus $H$ (left) and $H$
$^{2}$ (right) for a pure YBCO$_{6.6}$ crystal. In (b) the data
depart from the normal state $H^{2}$ dependence (dashed lines) below
a threshold fied $H_{c}^{^{\prime}}$ (arrows). In (c) the arrow
indicates the determination of $"H_{c2}"$ if a linear extrapolation
is used as often done \cite{mackenzie, Osofsky}, which leads to a
very small value of $\thicksim20$T.}
 \label{Fig.1}
\end{figure}

Above $H_{c}^{\prime}$ the linear fits of $\delta\rho$ vs $H^{2}$
indicated as broken lines in Fig.1b allow us to estimate the zero
field normal state resistivity from the zero field intercept
$\delta\rho_{0}(T)$ as $\rho_{n}(T)=\rho(0,T)+\delta\rho_{0}(T) $
and the normal state MR coefficient as
$a_{trans}(T)=\delta\rho_{n}(T)/\rho_{n}(T)H^{2}$. The $\rho_{n}(T)$
data displayed in Fig.2 can be well fitted down to 60K by a $T^{2}$
law : $\rho _{n}(T)=\rho_{0}+AT^{2}$ with
$\rho_{0}=38.4\mu\Omega.cm$ and $A=0.0066\mu\Omega.cm/K^{2}$. A
similar $T^{2}$ dependence of the resistivity has been reported in
various underdoped cuprates \cite{Ando3}.

\begin{figure}
\centering
\includegraphics[width=8.0cm]{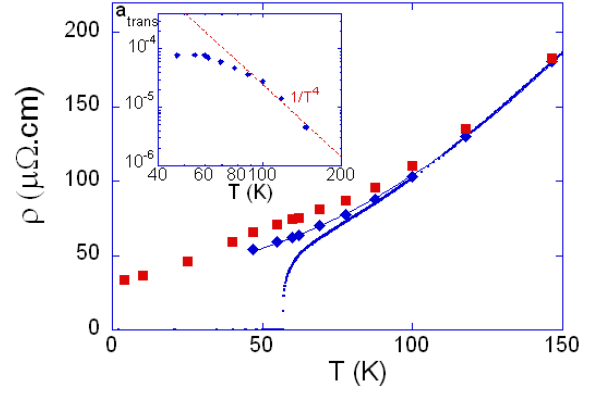}
\caption{(Color online)Temperature dependence of the resistivity for
pure YBCO$_{6.6}$. Zero field data are represented by the nearly
continuous curve while squares are data taken at 51T. Diamonds
correspond to the normal state resistivity $\rho_{n}(0,T)$
extrapolated to zero field, which is fitted by a $T^{2}$ variation
(solid line. The inset shows the transverse MR coefficient
$a_{trans}$ vs $T$ in a log-log scale. The data follows a $T^{-4}$
dependence at high $T$.}
\label{Fig.2}%
\end{figure}

As seen in the inset of Fig.2 where $a_{trans}$ is plotted versus $T$ in a
log-log scale, the MR coefficient follows a $T^{-4}$ power law down to
$\thicksim80K$. This $T^{-4}$ dependence, which has been found for the orbital
MR for $T>100$K in underdoped and optimally doped YBCO \cite{Harris}, has been
referred as the \textquotedblright modified Kohler's rule\textquotedblright%
 $a_{orb}$ $\propto\tan^{2}\Theta_{H}$ where $\Theta_{H}$ is the Hall angle.
Although a quantitative comparison of our data with this expression
cannot be done here as we have only measured the transverse MR
\cite{footnote1}, our observation of a $T^{-4}$ dependence confirms
that we do measure the normal state MR above $H_{c}^{^{\prime}}$.
Below $80K$ $a_{trans}$ saturates as clearly seen also in Fig.1d. A
saturation of $a_{orb}$ has been observed in LSCO at low $T$
\cite{Harris} and has been interpreted as the contribution of a
large impurity scattering term in this system. One also might expect
that the contribution of the residual $\rho_{0}$ becomes dominant at
low $T$ in our YBCO$_{6.6}$ crystal. Another possible explanation of
the saturation of the MR might come from the breakdown of the weak
field limit as $T$ (and thus $\rho_{n}$) decreases \cite{Tyler}.
Indeed, taking a hole
doping of $0.09$ per Cu in the CuO$_{2}$ plane gives $\omega_{c}%
\tau\thickapprox1$ at 60T for $\rho\thickapprox40\mu\Omega.cm$.

Although it is not possible to determine accurately $\rho_{n}(T)$\
and $H_{c}^{^{\prime}}$  at low $T$, the values of $\rho$ taken at
51T displayed in Fig.2 decrease with $T$ down to 4.2K. This behavior
is confirmed on another sample on which measurements have been done
down to 1.6K, which evidences that YBCO$_{6.6}$ displays a metallic
behavior in contrast to underdoped \textquotedblright low
$T_{c}$\textquotedblright\ cuprates for which upturns of the
resistivity were found at low $T$ \cite{Boebinger,Ono}

Our approach provides a reliable way to determine
$H_{c}^{^{\prime}}$ by distinguishing the normal state resistivity
from the superconducting contribution above 60K, but would become
ambiguous below, as the normal state is obviously not fully
accessible with such limited fields. In order to get insight into
the shape of the $H_{c}^{^{\prime}}$ line below $T_{c}$, samples
with lower $H_{c}^{^{\prime}}$ were required. We have attempted to
do so, by depressing $T_{c}$ with low $T$ electron irradiation
\cite{RA2}. The resistivity data are reported in Fig.3 for an
irradiated sample with $T_{c}=6.8K$. In that sample the large
resistivity ($>400\mu\Omega.cm$) yields $\omega_{c}\tau<0.1$, which
ensures that the weak field limit is always valid. From 30 to 100K
the $\delta\rho$ vs $H$ curves display a clear upward curvature,
which allows us to evidence unambiguously the normal state $H^{2}$
term and to determine $H_{c}^{^{\prime}}(T)$. Below 30K the
superconducting contribution is quite large but saturates rapidly in
field, as can be seen in Fig.3c and 3d from the parallel behavior of
the resistivity curves at high field. By continuity with the high
$T$ analysis, $H_{c}^{^{\prime}}(T)$ can be defined as shown in
fig.3d \cite{footnote2}.

\begin{figure}
\centering
\includegraphics[width=8.0cm]{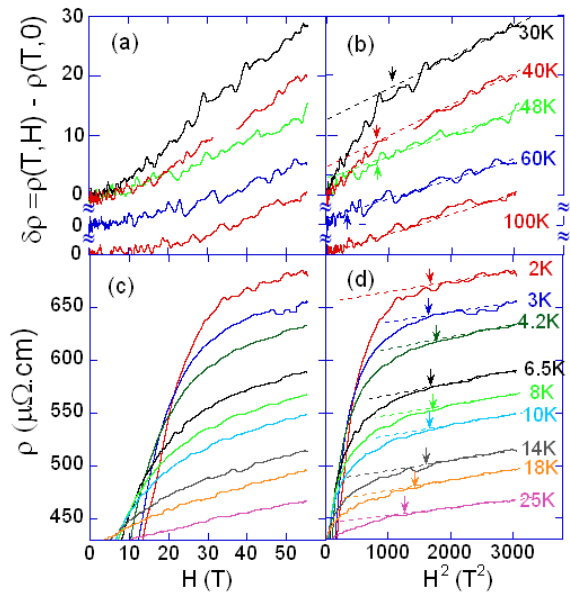}
\caption{(Color on line) Magnetoresistance above 30K and resistivity
below 30K are respectively plotted versus $H$ (left) and $H$ $^{2}$
(right). The data depart from the normal state $H^{2}$ dependence
(dashed lines) below a threshold fied $H_{c}^{^{\prime}}$ (arrows).
The upturn of resistivity at low $T$ is induced by disorder
\cite{RA1}.}%
\label{Fig.3}%
\end{figure}

The $T$ dependences of $H_{c}^{^{\prime}}$ deduced from this
analysis  are reported for the two samples in Fig.4 altogether with
data for two other irradiated samples (which are less accurate as
measurements have been performed using short field pulses). For all
the irradiated samples studied, the $H_{c}^{^{\prime}}$ line is well
fitted by a simple quadratic formula
$H_{c}^{^{\prime}}(T)=H_{c}^{^{\prime}}(0)\left[  1-(T/T_{c}^{^{\prime}}%
)^{2}\right]$. This legitimates the use of such a fit for the pure
sample for which low $T$ data were not accessible. These fits yield
$H_{c}^{^{\prime}}(0)=65(\pm3)$, $50(\pm3)$, $41(\pm3)$, $34(\pm4)$T
and $T_{c}^{^{\prime}}=$ $108(\pm3)$, $86(\pm3)$, $81(\pm3)$,
$75(\pm4)$K respectively for the different samples with decreasing
$T_{c}=57K$, $25K$, $6.8K$ and $3.5$K.

\begin{figure}
\centering
\includegraphics[width=8.0cm]{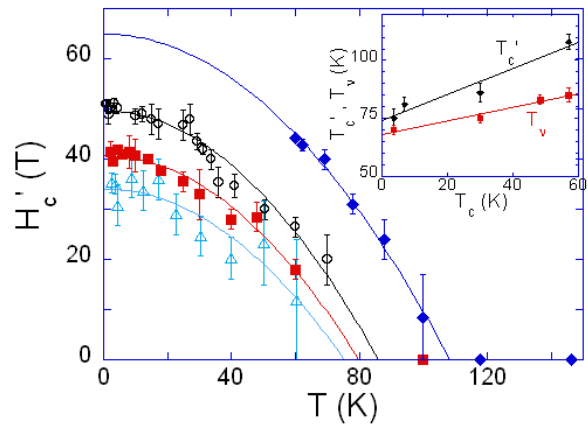}
\caption{(Color online)Variation of
$H_{c}^{^{\prime}}$ vs $T$ for YBCO$_{6.6}$ samples : pure
(diamonds) and irradiated ones with respective $T_{c}$ of 25
(circles), 6.8 (square) and 3.5K (triangles). The solid lines
represent the quadratic fits given in the text, which define
$T_{c}^{^{\prime}}$. The inset shows $T_{c}^{^{\prime}}$ and
$T_{\nu}$ versus $T_{c}$ for various samples.}
 \label{Fig.4}
\end{figure}

The experimental observation of a $H_{c}^{^{\prime}}(T)$ line which
terminates at $T_{c}^{^{\prime}}$ well above $T_{c}$ with a
quadratic $T$ dependence is strikingly similar to that expected for
the mean field crossover above the 2D Kosterlitz-Thouless (KT)
transition \cite{Doniach}. In such an approach the phase coherence
is destroyed at $T_{c}$ by a proliferation of thermal vortices,
while $H_{c}^{^{\prime}}(T)$ represents the line above which local
superconducting pairing vanishes. So our results share a lot of
similarities with those deduced from the anomalous Nernst effect
which has been interpreted within this vortex fluid picture
\cite{Wang-PRB2}. We can then directly compare $T_{c}^{^{\prime}}$
with the onset $T_{\nu}$ for the Nernst signal measured previously
on the same samples \cite{RA3}. One can see in the inset of Fig.4
that $T_{c}^{^{\prime}}$ is systematically found to be higher than
$T_{\nu}$, so that the MR measurements are sensitive to fluctuations
above the vortex regime (such as amplitude fluctuations).

Let us emphasize that the lack of detection of a flux flow
contribution to the conductivity has been often taken as detrimental
to the vortex scenario, although it has been argued that it could be
negligible \cite{Ioffe}. In the two fluid model the conductivity is
the sum of the superfluid $\sigma_{s}$ and the normal quasiparticle
$\sigma_{n}$ conductivities. Our experimental result allows us to
estimate a small but sizable $\sigma_{s}$ $\simeq0.1\sigma_{n}$ at
$T_{c}$ and $20$T in the pure sample. This small superconducting
contribution to the magnetoconductivity has been misregarded so far
in most attempts to determine $H_{c2}$ below $T_{c}$, the residual
variation of $\rho$ in high field being attributed to the normal
state. This ambiguity had been indeed noticed by some authors
\cite{resistiveHc2}.

One important aspect of our experimental results is that it allows us to study
the influence of disorder on the phase diagram. As already observed by Nernst
measurements for $T_{\nu}$ \cite{RA3}, the introduction of defects
considerably expands the range of superconducting fluctuations since $T_{c}$
decreases whereas we find here that $T_{c}^{^{\prime}}$ and $H_{c}^{^{\prime}%
}$ are only slightly depressed. In the most irradiated sample with $T_{c}%
=3.5$K, the $H_{c}^{^{\prime}}$ line terminates at $T_{c}^{^{\prime}}%
\simeq75K$, i.e. $\simeq20T_{c}$, while $H_{c}^{^{\prime}}$ is only
reduced by a factor 2. This weak variation of $H_{c}^{^{\prime}}$
which is related to the pairing strength in the KT approach, appears
coherent with the absence of variation of the superconducting
coherence peak at the antinodal direction probed by ARPES in
electron irradiated optimally doped Bi2212 \cite{Vobornik}. In the
KT approach the value of $T_{c}$ is directly related to the value of
the local phase stiffness at $T_{c}$ which is depressed from its 0K
value by thermal excitation of quasiparticles \cite{Lee}. As in
plane defects, such as Zn substitutions are well known to decrease
markedly the $T=0$ phase stiffness \cite{Nachumi}, this naturally
gives an explanation of the contribution of phase decoherence to the
observed quasi linear decrease of $T_{c}$ \cite{RA2}.

It seems legitimate to assimilate $H_{c}^{^{\prime}}$ with the upper
critical field $H_{c2}$ inferred from Nernst or magnetization
measurements \cite{Wang-science, Wang-PRL2}. An estimate of $H_{c2}$
\textit{below }$T_{c}$ has been taken as the extrapolated field for
which the Nernst signal should vanish \cite{Wang-science}.
Experiments above $T_{c}$ are hardly possible, the
Nernst signal becoming then very small. In our experiments, the $H_{c}%
^{^{\prime}}(T)$ line is most directly accessible \textit{above }$T_{c}$. We
observe in the most irradiated samples that $H_{c}^{^{\prime}}$ only begins to
decrease for $T>>T_{c}$. So our results on samples with low $T_{c}$ agree with
those obtained on families of \textquotedblright low $T_{c}$\textquotedblright%
\ cuprates where $H_{c2}$ was found nearly constant through $T_{c}$
\cite{Wang-science}. On the contrary, for the pure sample, we show that
$H_{c}^{^{\prime}}$ starts to decrease below $T_{c}$ thereby vanishing at
$T_{c}^{\prime}\thickapprox2T_{c}$. In optimally doped Bi2212 a similar
decrease of $H_{c2}$ from $\approx200$T at $35$K to $\approx90$T at $T_{c}%
=86$K has been inferred from diamagnetism data \cite{Wang-PRL2}. To
clarify why $H_{c}^{^{\prime}}(0)\ $is much smaller than 200T in
YBCO$_{6.6}$ we have analysed preliminary resistivity data taken on
an YBCO$_{7}$ crystal in which $T_{c}$ was depressed to 30K by
electron irradiation. Even in that case it was impossible to restore
the normal state below 35K with a magnetic field of 55T, resulting
in $H_{c}^{^{\prime}}(0)$ $>$70-80T. This ascertains that in the
pure compounds $H_{c}^{^{\prime}}(0)$ is much larger in YBCO$_{7}$
than in YBCO$_{6.6}$. So,the variation of $H_{c}^{^{\prime}}(0)$
from the underdoped case to the optimally doped one parallels the
variation of $T_{c}$ rather than that of the pseudogap which is
known to increase markedly upon underdoping.

In conclusion we have presented here a powerful method based on
resistivity measurements which allowed us to measure directly the
$T$ dependence of the mean field $H_{c2}$ in the largest domain
where superconductivity is experimentally detectable, at least as
fluctuations. The decrease of the phase stiffness induced by the
defects naturally yields the subsequent depression of the KT
transition temperature that is of $T_{c}$. Furthermore $H_{c2},
T_{c}^{^{\prime}}$ and $T_{\nu}$ which are related to pair formation
are depressed, although moderately, by the defects, in contrast to
the pseudogap $T^{\ast}$ which has been found insensitive to
disorder \cite{Alloul91}. So all experimental evidences obtained so
far in YBCO indicate that $T^{\ast}$ is not directly related to the
pairing energy scale. This remark sustains the proposal done by P.
Lee et al \cite{Lee} for an experimental confusion between two types
of pseudogaps, one linked to a survival above $T_{c}$ of the actual
SC gap and the other one initially detected by NMR, and associated
with singlet formation.

The experiments at the LNCMP have been financed by the contract FP6
"Structuring the European Research Area, Research Infrastructure
Action" contract R113-CT2004-506239.\ FRA and HA acknowledge the
NHMFL for financial support during their visit to NHMFL and thank F.
Balakirev for his help during the experiments and for his critical
reading of the manuscript.


\begin{thebibliography}{99}                                                                                               %
\bibitem {Boebinger}G.S.\ Boebinger et al., Phys.\ Rev.\ Lett. \textbf{77},
5417 (1996).

\bibitem {Ono}Ono et al., Phys.\ Rev.\ Lett. \textbf{85}, 638 (2000).

\bibitem {mackenzie}A. P. Mackenzie etal., Phys. Rev. Lett. \textbf{71}, 1238 (1993).

\bibitem {Osofsky}M.S.\ Osofsky et al., Phys. Rev. Lett. \textbf{71}, 2315 (1993).

\bibitem {Ando2}Y.\ Ando and K.\ Segawa, Phys. Rev. Lett. \textbf{88}, 167005 (2002).

\bibitem {Bouquet}F.\ Bouquet et al., Phys. Rev. B \textbf{74}, 064513 (2006).

\bibitem {Emery}V.J.\ Emery and S.A.\ Kivelson, Nature \textbf{374}, 434 (1995).

\bibitem {Corson}J.\ Corson et al., Nature \textbf{398}, 221 (1999).

\bibitem {Wang-PRB1}Y.\ Wang et al., Phys.\ Rev.\ B \textbf{64}, 224519-1 (2001).

\bibitem {RA3}F.Rullier-Albenque et al.,Phys.\ Rev.\ Lett. \textbf{96}, 067002 (2006).

\bibitem {Wang-PRL2}Y.\ Wang et al., Phys.\ Rev.\ Lett. \textbf{95}, 247002 (2005).

\bibitem {Wang-science}Y.\ Wang et al., Science \textbf{299},86 (2003).

\bibitem {Doniach}S.\ Doniach and B.A.\ Huberman, Phys.\ Rev.\ Lett.
\textbf{42}, 1169 (1979).

\bibitem {Harris}J.M.\ Harris et al., Phys.\ Rev.\ Lett. \textbf{75}, 1391 (1995).

\bibitem {Ando3}Y.\ Ando et al, Phys.\ Rev.\ Lett. \textbf{92},
197001\textbf{\ }(2004).

\bibitem {footnote1}The orbital part of the MT is obtained by $a_{orb}%
=a_{trans}-a_{long}$.

\bibitem {Tyler}A.W.\ Tyler et al., Phys.\ Rev.\ B \textbf{57}, R728 (1998).

\bibitem {footnote2}One can notice here that using Fig.3c to point
$H_{c}^{^{\prime}}$ as the field at which the curves deviate from the high
field linear behavior, as sometimes done, would give nearly identical values
of $H_{c}^{^{\prime}}$ for $T\leqslant14$K.

\bibitem {RA1}F.Rullier-Albenque, H.\ Alloul, R.\ Tourbot, Phys.\ Rev.\ Lett.
\textbf{87},157001 (2001).

\bibitem {Wang-PRB2}Y.\ Wang, L.\ Li, N.P.\ Ong, Phys.\ Rev.\ B \textbf{73},
024510 (2006).

\bibitem {Ioffe}L.B.\ Ioffe and A.J.\ Millis, Phys.\ Rev.\ B \textbf{66},
094513 (2002).

\bibitem {resistiveHc2}Y.\ Ando et al., Phys.\ Rev.\ B \textbf{60}, 12475
(1999), P Fournier et R. Greene, ibid. \textbf{68} 094507 (2003).

\bibitem {Nachumi}B.\ Nachumi et al., Phys.\ Rev. Lett. \textbf{77}, 5421 (1996).

\bibitem {Vobornik}I.\ Vobornik et al., Phys.\ Rev.\ Lett. \textbf{82}, 3128 (1999).

\bibitem {Lee}P.A.\ Lee, N.\ Nagaosa, X.G.\ Wen,
Rev.\ Mod.\ Phys.\ \textbf{78}, 17 (2006).

\bibitem {RA2}F.Rullier-Albenque, H.\ Alloul, R.\ Tourbot, Phys.\ Rev.\ Lett.
\textbf{91},047001 (2003).

\bibitem {Alloul91}H.\ Alloul et al., Phys.\ Rev.\ Lett. \textbf{67}, 3140 (1991).
\end{thebibliography}
\end{document}